\documentclass[amssymb,prb,twocolumn,floats,amsmath,showpacs,superscriptaddress]{revtex4}
\usepackage{bm}
\usepackage{graphicx}

\begin{document}


\title{Influence of  $s,p-d$ and $s-p$ exchange couplings on exciton
splitting in Zn$_{1\textrm{-}x}$Mn$_{x}$O}


\author{W.~Pacuski}
\email{Wojciech.Pacuski@fuw.edu.pl}
\affiliation{Faculty of Physics, University of Warsaw,
Ho\.za 69, PL-00-681 Warszawa, Poland}
\affiliation{Institut N\'eel/CNRS-Universit\'e J. Fourier,
Bo\^{\i}te Postale 166, F-38042 Grenoble, France}

\author{J.~Suf\mbox{}fczy\'{n}ski}
\affiliation{Faculty of Physics, University of Warsaw,
Ho\.za 69, PL-00-681 Warszawa, Poland}

\author{P. Osewski}
\affiliation{Faculty of Physics, University of Warsaw,
Ho\.za 69, PL-00-681 Warszawa, Poland}

\author{P.~Kossacki}
\affiliation{Faculty of Physics, University of Warsaw,
Ho\.za 69, PL-00-681 Warszawa, Poland}

\author{A.~Golnik}
\affiliation{Faculty of Physics, University of Warsaw,
Ho\.za 69, PL-00-681 Warszawa, Poland}

\author{J.~A.~Gaj}
\thanks{Deceased.}
\affiliation{Faculty of Physics, University of Warsaw,
Ho\.za 69, PL-00-681 Warszawa, Poland}

\author{C.~Deparis}
\affiliation{Centre de Recherches sur l'H\'et\'ero\'epitaxie  et ses
Applications, CNRS,\\rue Bernard Gr\'egory, Parc Sophia Antipolis,
F-06560 Valbonne, France}

\author {C.~Morhain}
\affiliation{Centre de Recherches sur l'H\'et\'ero\'epitaxie  et ses
Applications, CNRS,\\rue Bernard Gr\'egory, Parc Sophia Antipolis,
F-06560 Valbonne, France}

\author{E.~Chikoidze}
\affiliation{Groupe d'Etude de la Mati\`{e}re Condensée, CNRS-Université
de Versailles,
45, Ave Etats Unis, 78035, Versailles, France}

\author{Y. Dumont}
\affiliation{Groupe d'Etude de la Mati\`{e}re Condensée, CNRS-Université
de Versailles,
45, Ave Etats Unis, 78035, Versailles, France}

\author{D.~Ferrand}
\affiliation{Institut N\'eel/CNRS-Universit\'e J. Fourier,
Bo\^{\i}te Postale 166, F-38042 Grenoble, France}

\author{J.~Cibert}
\affiliation{Institut N\'eel/CNRS-Universit\'e J. Fourier,
Bo\^{\i}te Postale 166, F-38042 Grenoble, France}

\author{ T.~Dietl}
\affiliation{Institute of Physics, Polish Academy of Sciences, al.
Lotnik\'{o}w 32/46, PL-02-668 Warszawa, Poland}
\affiliation{Faculty of Physics, University of Warsaw,
Ho\.za 69, PL-00-681 Warszawa, Poland}


\begin{abstract}

This work presents results of near-band gap magnetooptical studies on
Zn$_{1\textrm{-}x}$Mn$_{x}$O epitaxial layers. We observe excitonic
transitions in reflectivity and photoluminescence, that shift towards
higher energies when the Mn concentration increases and split nonlinearly under the magnetic
field. Excitonic shifts are determined by the $s,p-d$ exchange coupling
to magnetic ions, by the electron-hole $s-p$ exchange, and the spin-orbit
interactions. A quantitative description of the magnetoreflectivity
findings indicates that the free excitons $A$ and $B$ are associated with the $\Gamma_7$ and $\Gamma_9$
valence bands, respectively, the order reversed as compared to wurtzite GaN.
 Furthermore, our results show that the magnitude of the
giant exciton splittings, specific to dilute magnetic semiconductors, is
unusual: the magnetoreflectivity data is described by an effective
exchange energy $N_0(\beta^{\mathrm{(app)}}-
\alpha^{\mathrm{(app)}})=+0.2\pm 0.1$~eV, what points to small and positive $N_0\beta^{\mathrm{(app)}}$. It is shown that both the
increase of the gap with $x$ and the small positive value of the
exchange energy $N_0\beta^{\mathrm{(app)}}$  corroborate recent theory
describing the exchange splitting of the valence band in a non-perturbative
way, suitable for the case of a strong $p-d$ hybridization.
\end{abstract}



\maketitle


\section {Introduction}

It is increasingly clear\cite{Dietl:2010_NM} that the understanding and
functionalization of dilute magnetic semiconductors (DMSs) and dilute
magnetic oxides (DMOs) has to proceed {\em via} detailed investigations of
spin-dependent couplings between effective mass carriers residing in $sp$
bands and electrons localized on $d$ shells of transition metals. According
to a recent model,\cite{Dietl:2008_PRB} a local potential produced by a
strong $p-d$ hybridization specific to wide bandgap DMSs and DMOs may give
rise to the presence of a hole bound state in the gap, even if the
transition metal (TM) ion acts as an isoelectronic impurity. In this strong
coupling case, the standard theory, invoking virtual crystal and molecular
field approximations to describe exchange splitting of bands, ceases to be
valid.\cite{Dietl:2008_PRB} A non-perturbative treatment of the $p-d$
coupling within the generalized alloy theory\cite{Tworzydlo:1994_PRB}
demonstrated that the spin splitting of {\em extended} valence band states
resulting from interactions with randomly distributed magnetic impurities,
each acting as a deep hole trap, is reversed and much
reduced.\cite{Dietl:2008_PRB} This view was recently supported by {\em ab
initio} computations within an LSDA + U approach for a series of (II,Mn)VI
compounds.\cite{Chanier:2009_PRB} It was also
suggested\cite{Pacuski:2008_PRL} that the developed theoretical
framework\cite{Dietl:2008_PRB} explains an anti-crossing-like behavior of
bands in mismatched alloys, such as GaAs$_{1\textrm{-}x}$N$_{x}$, often
assigned to the presence of resonant impurity states.\cite{Wu:2002_a}

The predictions of the strong coupling model\cite{Dietl:2008_PRB} were
experimentally confirmed for wurtzite  Ga$_{1\textrm{-}x}$Fe$_{x}$N and
Ga$_{1\textrm{-}x}$Mn$_{x}$N, where a small and positive value of the
apparent $p-d$ exchange energy $N_0\beta^{\text{(app)}}$ was inferred from
magneto-reflectivity studies of free
excitons.\cite{Pacuski:2008_PRL,Pacuski:2007_PRB,Suffczynski:2011_PRB} An unexpectedly small
magnitude of $N_0\beta^{\text{(app)}}$ was also found in
Zn$_{1\textrm{-}x}$Co$_{x}$O,\cite{Pacuski:2006_PRB} whose sign depended on
the assumed ordering of the valence bands on ZnO, a much disputed issue in
the physics of this important compound.\cite{Thomas:1960,Reynolds:1999_PRB, Gil:2001_PRB,Lambrecht:2002_a,Wagner:2009_PRB_b,Ding:2009_JAP}

The validity of the model\cite{Dietl:2008_PRB} for the description of
Zn$_{1\textrm{-}x}$Mn$_{x}$O has not yet been clarified. Recent
transmission experiments indicated the presence of an absorption band in
the sub-gap spectral region.\cite{Godlewski:2010_OM,Johnson:2010_PRB} In
agreement with the strong coupling theory, this additional absorption was
considered to originate from photoexcitation of a hole in a Mn bound state
and an electron in the conduction band. The corresponding optical
cross-section was found to be large enough to shadow internal Mn$^{2+}$
transitions, expected to appear within the same spectral region.  The
trapping of holes by Mn ions explained also a decay of photoluminescence
(PL) intensity with the Mn concentration.

The previous PL study\cite{Przezdziecka:2006_SSC} of bound excitons in
Zn$_{1\textrm{-}x}$Mn$_{x}$O under an applied magnetic field indicated a much
smaller magnitude and a reverse sign of the splitting compared to other
II-VI DMSs, in qualitative agreement with the
theory.\cite{Dietl:2008_PRB} This view was called into question by results
of magnetic circular dichroism (MCD) measurements which indicated that the
sign and an overall magnitude of the MCD signal were actually similar to
other II-VI DMSs.\cite{Johnson:2010_PRB}  Moreover, it can be argued that
conclusions based on PL results are misleading if magnetic ions act as PL
killers. In such a case, if the sample is not perfectly homogenous, the
emission comes primarily from regions with low magnetic ion concentrations
and, thus, characterized by a relatively small magnitude of the spin
splitting.  This problem is not important in the case of reflectivity,
whose amplitude is virtually independent of the magnetic ion concentration.

We have, therefore, decided to perform, in addition to the PL and the MCD
measurements, reflectivity studies on high quality samples exhibiting sharp
excitonic lines. Extensive magnetoreflectivity results and their
quantitative analysis taking into account excited states as well as both
$s,p-d$ and $s-p$ (electron-hole) exchange interactions reported here
suggest that the excitons $A$ and $B$ correspond to different valence bands
($\Gamma_7$ and $\Gamma_9$, respectively) than in GaN. Furthermore, the findings point to the positive sign and to a
reduced magnitude of $\beta^{\text{(app)}}$, confirming the applicability
of the strong coupling model\cite{Dietl:2008_PRB} for ZnO-based DMSs. We
explain mutually opposite polarization of the lowest excitonic lines seen in reflectivity and photoluminescence measurements.


\section {Samples and experimental}

A series of about 1~$\mu$m thick Zn$_{1\textrm{-}x}$Mn$_{x}$O layers has been
grown by molecular beam epitaxy (MBE) and metal-organic vapor phase epitaxy
(MOVPE) on sapphire substrates.  The $c$-axis is parallel to the growth
direction and the Mn concentrations $x$, determined by secondary ion mass
spectroscopy (SIMS), is 0.14, 0.6, 1.4, and 2.6\% for the layers grown
by MBE, and 2.5\%  for the film obtained by MOVPE.
A high pressure Xe lamp was employed for reflectivity and transmission
experiments. Photoluminescence was excited by a He-Cd laser. Circularly
polarized light was analyzed by an achromatic quarter-wave plate and fused
silica linear polarizer. A Peltier cooled CCD camera coupled to a 1800
gr/mm grating monochromator was used to record the spectra. The magnetooptical data were
collected in the Faraday geometry and with the light incidence normal to
the film plane. The sign of circular polarization $\sigma^{\pm}$ was checked
by a reference measurement of the giant Zeeman splitting of excitons in
Cd$_{1\textrm{-}x}$Mn$_{x}$Te.


\section {Results}

\subsection {Zero field spectroscopy}

\begin{figure}
\includegraphics[width=1\linewidth]{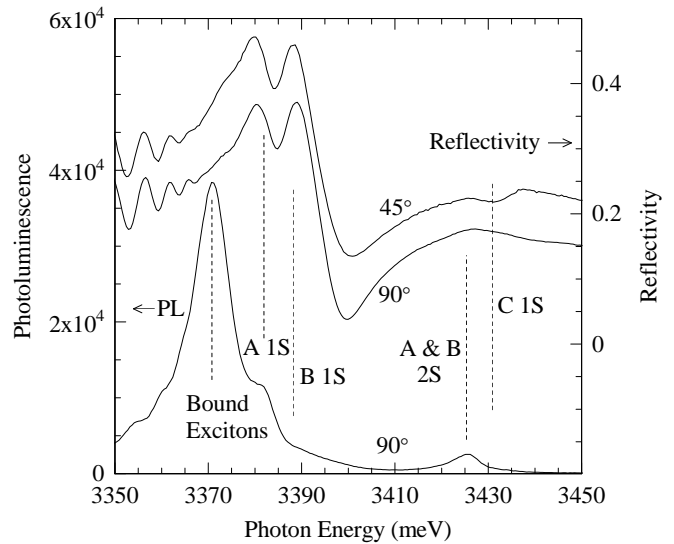}
\centering \caption{Reflectivity and photoluminescence of Zn$_{1\textrm{-}x}$Mn$_{x}$O
with $x=0.6$\% at $T=1.6$~K. The reflectivity is measured with incidence angle of
45~degrees (upper spectrum shifted for clarity) or with
normal incidence (middle spectrum). Estimated
spectral positions of excitonic lines are indicated.
} \label{fig:znmno_lines}
\end{figure} 

Reflectivity and photoluminescence spectra of diluted
Zn$_{1\textrm{-}x}$Mn$_{x}$O ($x=0.6$\%) are shown in
Fig.~\ref{fig:znmno_lines}. Several features related to excitonic
transitions are seen. In general, three kinds of excitons associated with
three closely lying valence bands,  are expected in wurtzite
semiconductors. They are traditionally refereed to as $A$, $B$, and $C$,
where the ground state of the exciton $A$ has the lowest and of the exciton
$C$ the highest photoexcitation energy. Accordingly, in the reflectivity spectrum collected for the incidence angle of 45 degrees all three 1S excitons $A$, $B$, and $C$ are observed (Fig.~\ref{fig:znmno_lines}, topmost spectrum). The $A-B$ splitting is about 7 meV, the $B-C$ splitting is about 45 meV. Due to a weak spin-orbit coupling and optical selection rules no exciton $C$ is observed in the reflectivity spectra obtained with the normal incidence (Fig.~\ref{fig:znmno_lines}, middle spectrum). The 2S excitons are much weaker than the 1S excitons and are shifted about 40 meV to higher energies.
In the PL spectrum of the same sample (Fig.~\ref{fig:znmno_lines}, bottom spectrum), the dominating line, identified as
the recombination of a donor bound exciton, is shifted towards lower energies by 11 meV as compared to a weaker line of free exciton $A$ (1S). Emission from the 2S exited states is well
visible about 40 meV above the free excitons. Comparing to a reference ZnO sample (not shown) excitonic transitions in $x=0.6$\% sample are shifted by 5 meV towards higher energies.

\begin{figure}
\includegraphics[width=0.9\linewidth]{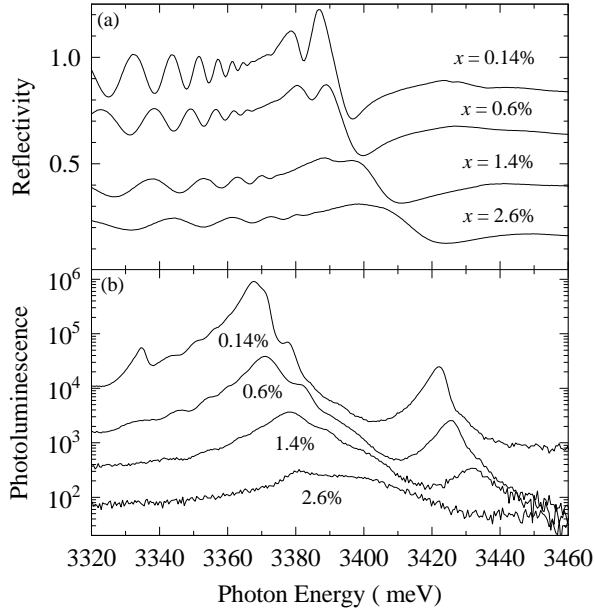}
\centering \caption{Reflectivity (a) and photoluminescence (b) spectra
collected at 1.6~K for Zn$_{1\textrm{-}x}$Mn$_{x}$O with Mn concentrations
0.14, 0.6, 1.4, and 2.6\%. Reflectivity spectra are shifted vertically for clarity. Intensity of photoluminescence spectra decreases by orders of magnitude with increasing the Mn concentration.} \label{fig:0znmno}
\end{figure}

Figures~\ref{fig:0znmno}(a) and \ref{fig:0znmno}(b) show the reflectivity
and PL spectra, respectively for samples with different  Mn  concentrations
$x$. Optical transitions of the $A$ and $B$ excitons are well resolved in
the reflectivity spectra for $x$ up to $1.4$\%. With increasing $x$ the
excitonic transitions broaden and shift towards higher energies, in qualitative consistency with previous reports.\cite{Johnson:2010_PRB} The same
trend is observed for transitions seen in PL (Fig.~\ref{fig:0znmno}(b)). 

\begin{figure}[h]
\includegraphics[width=0.7\linewidth,clip]{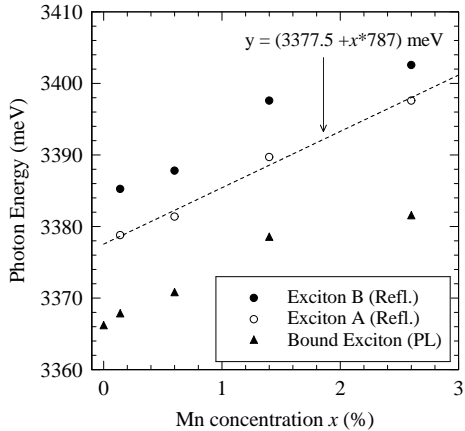}
\centering \caption{Exciton energies determined from the spectra shown in
Fig.~\ref{fig:0znmno}, as a function of the Mn concentration $x$ in
Zn$_{1\textrm{-}x}$Mn$_{x}$O.
Empty circles -- free exciton $A$ (reflectivity), full circles -- free exciton B (reflectivity), triangles -- bound
exciton (main PL line). The dashed line is a linear fit to the energy of
the free exciton $A$.} \label{fig:znmno_eg}
\end{figure}

The fitting of the reflectivity spectra by the polariton
model\cite{Pacuski:2008_PRL,Pacuski:2006_PRB} provides the energy values of the
$A$ and $B$ excitons, even in the case of the samples with the highest
concentration of magnetic ions $x=2.6$\%, where relevant lines are
rather broad. Figure~\ref{fig:znmno_eg} shows the energy positions of the $A$
and $B$ excitons determined from the spectra presented in
Fig.~\ref{fig:0znmno}(a). The energy of the bound exciton determined from
the PL spectra of  Fig.~\ref{fig:0znmno}(b) is also shown. A linear
dependence of the energy band gap on $x$ confirms previous reports on
Zn$_{1\textrm{-}x}$Mn$_{x}$O,\cite{Fukumura:1999_a, Godlewski:2010_OM, Johnson:2010_PRB,Chikoidze:2007_JAP}
 and is consistent with theoretical expectations for the case of the strong
$p-d$ hybridization.\cite{Dietl:2008_PRB} Similar slope for reflectivity and PL (Fig.~\ref{fig:znmno_eg}) shows that regions with lower Mn concentration do not dominate the PL spectrum.

\subsection {Magneto-reflectivity}
\label{Magneto-reflectivity}

\begin{figure}
\includegraphics[width=\linewidth,clip]{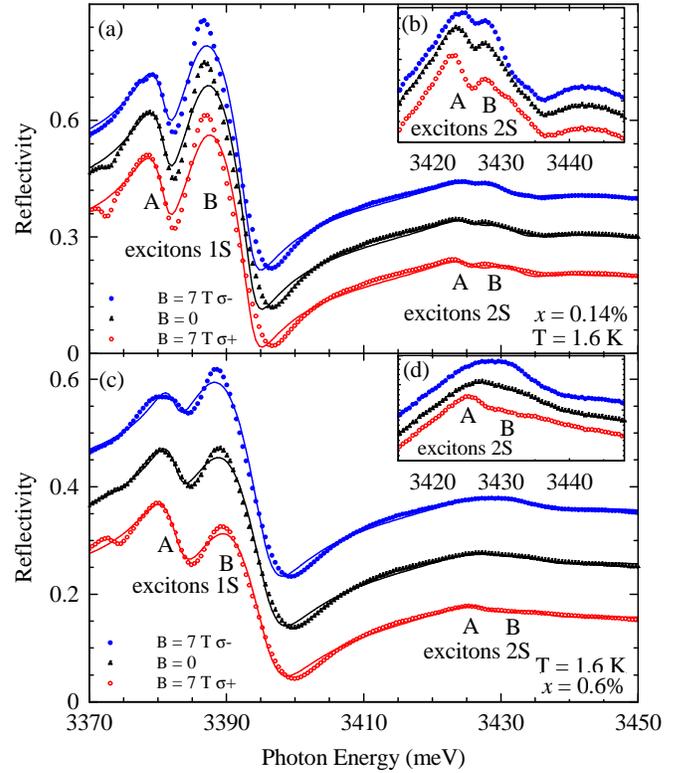}
\centering
\caption {(color online) Reflectivity spectra (symbols) of
Zn$_{1-x}$Mn$_{x}$O with $x=0.14$\% (a,b) and $x=0.6$\% (c,d). Lines show
fits by the polariton model. Features related to 1S and 2S excitons $A$
and $B$ are marked.
Spectral region of 2S excitons is plotted in a zoomed Y
scale in the insets (b) and (d). Full (empty) circles are related to measurement at
$\sigma^-$ ($\sigma^+ $) circular polarization and magnetic field $B=7$~T,
triangles to measurement at $B=0$~T.
} \label{fig:diluted_spc}
\end{figure}

Figure~\ref{fig:diluted_spc} presents reflectivity spectra of samples with
$x=0.14$\% and $0.6$\% recorded in the magnetic field $B =0$ and 7~T for
two circular polarizations. A high optical quality of the samples allows us
to resolve transitions of 1S as well as of 2S excitonic states, as
indicated in the plot.

In order to determine the exact energy positions of excitons from
reflectivity spectra, we fit the data employing the polariton
model.\cite{Pacuski:2008_PRL,Pacuski:2006_PRB} The model takes into account
excitons $A$ and $B$, their excited states, and transitions to continuum of
states through a residual dielectric function. Since
the exciton $C$ is not observed, its oscillator strength is assumed to be equal to zero.
All excited states are
described by a single fitting parameter\cite{Pacuski:2008_PRL} ($\Gamma_{\infty}$). The importance
of the excited states for the interpretation of reflectivity spectra in GaN
was pointed out by St\mbox{\c{e}}pniewski {\em et al.}\cite{Stepniewski:2003_PRL} We find that the inclusion of excited states such as 2S is also crucial for a
satisfactory description of reflectivity spectra in the case of ZnO. Interestingly, as clearly evidenced in Fig.~\ref{fig:diluted_spc}(a,b) the energy splitting between 2S $A$ and $B$ excitons is smaller than between 1S $A$ and $B$ excitons. This reveals that the magnitude of the electron-hole exchange interaction is much larger in the case of 1S excitons.

\begin{figure}
\includegraphics[width=\linewidth,clip]{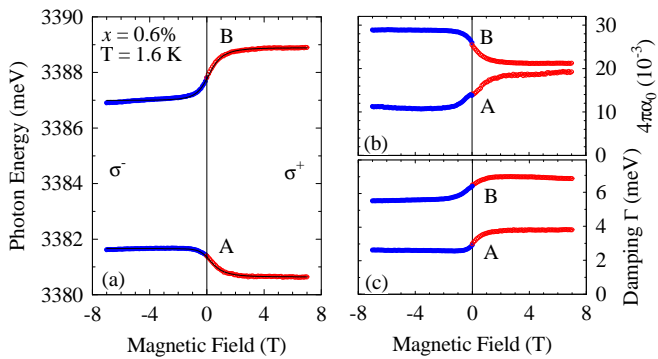}
\centering \caption[Spectral position of $A$ and $B$ excitons in
Zn$_{0.994}$Mn$_{0.006}$O]{(color online) Transition energies (a), polarizabilities (b), and
damping parameters (c) determined for $A$ and $B$ excitons in
Zn$_{0.994}$Mn$_{0.006}$O sample. Lines in (a) represent fit with excitonic
model, Brillouin function and  with parameters given in Table
\ref{tab:znmnoZ435}.
} \label{fig:diluted_position}
\end{figure}

\begin{figure}
\includegraphics[width=1\linewidth,clip]{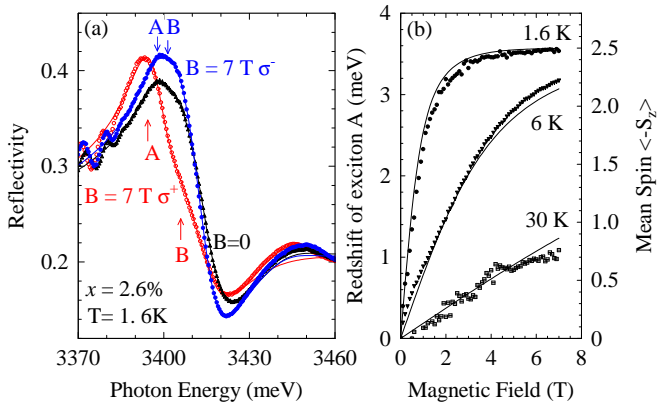}
\centering \caption[Polariton fit of the reflectivity of
Zn$_{1\textrm{-}x}$Mn$_{x}$O ($x=2.6$\%)]{(color online) (a) Reflectivity spectra
(symbols) of Zn$_{1-x}$Mn$_{x}$O with $x=2.6$\%, measured at $B=0$
and $B=7$~T in two circular polarizations at $T=1.6$~K. Solid lines show
fits with the polariton model. Arrows show the determined positions of
excitons. (b) Redshift of the exciton $A$ as a function of the magnetic
field $B$, as determined for in Zn$_{0.974}$Mn$_{0.026}$O at three
temperatures $T$ (left axis).  The redshift is compared to the mean spin of
Mn$^{2+}$ ions (right axis), given by the paramagnetic Brillouin function
B$_{5/2}(T,B)$.} \label{fig:high_Mn}
\end{figure}

Examples of experimental and calculated reflectivity spectra are shown in
Fig.~\ref{fig:diluted_spc}(a,c) and Fig.~\ref{fig:high_Mn}(a). Parameters
of the 1S excitonic transitions (transition energies, polarizabilities, and
damping parameters) obtained from the fit for the sample containing $0.6$\% of Mn are plotted as
a function of the magnetic field in Fig.~\ref{fig:diluted_position}. The analysis of the two circular polarizations of the light shows that both excitons $A$ and $B$ split in the magnetic field. However, in contrast
to Ga$_{1\textrm{-}x}$Fe$_{x}$N\cite{Pacuski:2008_PRL} and
Ga$_{1\textrm{-}x}$Mn$_{x}$N,\cite{Suffczynski:2011_PRB} the splitting
magnitude is larger in the case of the exciton $B$ (2~meV at saturation)
than in the case of the exciton $A$ (1~meV at saturation). Since a larger
splitting is expected for the $\Gamma_9$ band (derived from the heavy hole
$j = 3/2$ band) than for the $\Gamma_7$ case, we conclude that the $A$
exciton is associated with the $\Gamma_7$ band in ZnO, in contrast to GaN,
where the relevant band is $\Gamma_9$. We note that such an ordering has been proposed for ZnO by Thomas\cite{Thomas:1960} and
recently confirmed using the analysis of donor bound excitons in ZnO bulk material.\cite{Wagner:2009_PRB_b,Ding:2009_JAP}

According to the fitting results collected in
Fig.~\ref{fig:diluted_position}, the field-induced excitonic shifts are
much more pronounced in the $\sigma^+ $ circular polarization than in the
$\sigma^-$ one. This
stems from the presence of an electron-hole exchange interaction that
leads to a mixing between $A$ and $B$ excitonic
states.\cite{Pacuski:2006_PRB,Gil:2001_PRB} When the energy distance between excitons
$A$ and $B$ increases (in $\sigma^+$), exciton mixing becomes weaker. In
contrast, in the case of the $\sigma^- $ circular polarization, the energy
separation between excitons $A$ and $B$ decreases in the magnetic field.
This enhances the exciton mixing and the resulting anticrossing behavior, which
reduces strongly the  field-induced shift of the corresponding excitonic
states.

Once determined the exciton energies and their shifts in the magnetic field, we employ the theory of free excitons in wurtzite DMSs\cite{Pacuski:2008_PRL,Pacuski:2006_PRB,Pacuski:2010} in order to assess the
selection rules and to evaluate the values of relevant parameters for
Zn$_{1\textrm{-}x}$Mn$_{x}$O. As shown in Table~\ref{tab:znmnoZ435}, the fitting
procedure provides the magnitudes of the crystal field splitting
$\tilde\Delta_1$, spin-orbit parameters $\Delta_2$ and $\Delta_3$ as well as the electron-hole exchange energy $\gamma$ and a combination of the $s,p-d$ exchange energies, $N_0(\beta^{\mathrm{(app)}}\mathrm{-}\alpha^{\mathrm{(app)}})$, where $N_0$ is the cation concentration; $\beta^{\mathrm{(app)}}$ and
$\alpha^{\mathrm{(app)}}$ are the apparent $p-d$ and $s-d$ (respectively) exchange integrals describing the splitting of the extended hole and electron states which form the excitons. The presence of antiferromagnetic interactions between the
nearest neighbor Mn pairs\cite{Kolesnik:2006_JAP} is taken into account by
introducing an effective Mn concentration, $x_{\text{eff}} = x(1-x)^{12}$.

Parameter $\Delta_2$ describes the parallel component of spin-orbit interaction that induces the $A$-$B$ valence band splitting.
According to a common interpretation\cite{Lambrecht:2002_a,Pacuski:2006_PRB} a positive value of $\Delta_2$ would imply that the order of
the valence bands is the same in ZnO and GaN (namely, $\Gamma_9$, $\Gamma_7$, $\Gamma_7$).
However, in the case of the present study, the value of the $\Delta_2 = 0.5 \pm 1$ meV determined from the fit is very small. As a consequence, the perpendicular component of the spin orbit interaction ($\Delta_3$) governs the valence band ordering. Parameter $\Delta_3$ describes the $\Gamma_7$-$\Gamma_7$ valence bands interaction that pushes one of the $\Gamma_7$ bands to the top of the valence bands. In the view of the above considerations, the observed valence band ordering is $\Gamma_7$, $\Gamma_9$, $\Gamma_7$.

For excitons the situation is more complex. The electron-hole $s-p$ exchange interaction increases the splitting between excitons $A$ and $B$ as compared to valence band distance. Moreover, $s-p$ exchange interaction mixes valence bands within excitonic states. Consequently, the hole in the exciton $A$ originates partially from both $\Gamma_7$ and $\Gamma_9$ valence bands. However, the main contribution comes from the $\Gamma_7$ valence band, rather than from the $\Gamma_9$ band, as it would in GaN.  The difference in the exciton origin accounts also for a reverse sign and
amplitude of the giant Zeeman splitting of the excitons $A$ and $B$ in Zn$_{1\textrm{-}x}$Mn$_{x}$O and
Zn$_{1\textrm{-}x}$Co$_{x}$O,\cite{Pacuski:2006_PRB} compared to Ga$_{1\textrm{-}x}$Fe$_{x}$N and
Ga$_{1\textrm{-}x}$Mn$_{x}$N.\cite{Pacuski:2008_PRL,Suffczynski:2011_PRB}
The fitting procedure leads also to $N_0(\beta^{\mathrm{(app)}}-
\alpha^{\mathrm{(app)} })=+0.22 \pm 0.05 $~eV, where the main contribution to the error comes from the uncertainty in the $x$ value.

\begin{table}
\centering
\begin{tabular}{cccccccc}
\vspace{0.1cm}
$E_A$ &$\tilde\Delta_1$ & $\Delta_2$ & $\Delta_3$ & $\gamma$
& ${N_0(\beta^{\mathrm{(app)}}\mathrm{-}\alpha^{\mathrm{(app)}})}$ \\
\hline
\\[-0.2cm]
3381.4~~ & 51.1~~ & 0.5~~ & 11.6~~&
2.7~~ &$+220$   \\ [0.2cm]
\hline
\end{tabular}
\caption{
Energies in meV of the ground state exciton $A$, $E_A$; crystal field
splitting $\tilde\Delta_1$; spin-orbit parameters $\Delta_2$ and
$\Delta_3$;  $s-p$ and $s,p-d$ exchange interactions,  $\gamma$ and
$N_0(\beta^{\mathrm{(app)}}\mathrm{-}\alpha^{\mathrm{(app)}})$,
respectively describing energies of excitons $A$ and $B$ as a function of
the magnetic field in Zn$_{1\textrm{-}x}$Mn$_{x}$O with $x=0.6$\%.}
\label{tab:znmnoZ435}
\end{table}



Positions of $A$ and $B$ excitons obtained from the polariton fit for the
case of Zn$_{1\textrm{-}x}$Mn$_{x}$O with higher Mn content, $x=2.6$\%, are indicated by arrows
in Fig.~\ref{fig:high_Mn}.
When a magnetic field is applied a substantial redshift of the exciton $A$ in the $\sigma^+ $ polarization
is clearly visible even without the polariton fit. Since the low-energy
component of the exciton $A$ does not anticross with other excitons, the
influence of the electron-hole exchange interaction on its shift is weak.
Accordingly, as shown in Figure~\ref{fig:high_Mn}(b), the redshift of the
exciton $A$ is directly proportional to the paramagnetic Brillouin function
B$_{5/2}(T,B)$, where $T$ is the sample temperature. The data confirm a
dominant contribution of the $s,p-d$ exchange interaction to the giant
Zeeman splitting and indicate that the spin-spin coupling between next
nearest Mn neighbors is weak in Zn$_{1\textrm{-}x}$Mn$_{x}$O.

\subsection {Magnetic circular dichroism and photoluminescence}
\label{sec:znmnoPL}
\begin{figure}
\includegraphics[width=1\linewidth,clip]{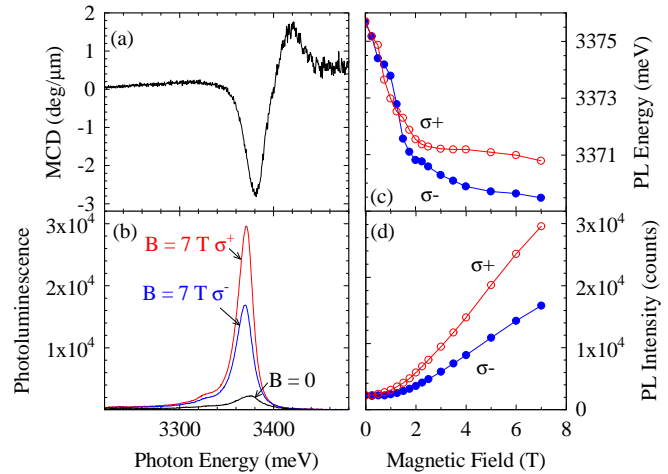}
\centering \caption[MCD and PL of Zn$_{1\textrm{-}x}$Mn$_{x}$O
($x=2.5$\%)] {(color online) Magnetic Circular Dichroism (MCD) in 6~T (a) and
photoluminescence (PL) spectra measured at B = 0 and B = 7 T
in two circular polarizations (b) for
Zn$_{1-x}$Mn$_{x}$O with 2.5\%~of Mn at T = 1.6~K. Spectral position (c) and
amplitude (d) of the PL peak vs. magnetic field.} \label{fig:znmno_PL}
\end{figure}

According to the magneto-reflectivity data discussed above, the redshifted
component of the lowest exciton line acquires the $\sigma^+$  polarization
in the presence of a magnetic field. This conclusion is consistent with the sign of MCD
signals reported for (Zn,Mn)O in the free exciton spectral
region.\cite{Johnson:2010_PRB} In contrast, the exciton line detected in PL
showed a redshift, but exhibited the  $\sigma^-$  polarization.\cite{Przezdziecka:2006_SSC}
In order to clarify this issue, we have carried out MCD and PL
measurements on the same Zn$_{0.975}$Mn$_{0.025}$O film.
The shape and the sign of the transmission MCD signal
(Fig.~\ref{fig:znmno_PL}a) are the same as reported previously
for (Zn,Mn)O,\cite{Johnson:2010_PRB,Ando:2001_JAP} and other II-VI DMSs,  including
Zn$_{1\textrm{-}x}$Co$_{x}$O.\cite{Pacuski:2006_PRB,Ando:2002_APL}

A comparison of Figs.~\ref{fig:znmno_PL}(a) and~\ref{fig:znmno_PL}(b)
indicates that the spectral position of the PL line is shifted towards
lower energies with respect to the MCD maximum. This can be expected as
transmission measurements probe primarily free excitons while PL involves
mainly excitons bound to residual impurities. The incorporation of magnetic
ions leads to a significant quenching of the excitonic PL intensity in
these systems, particularly if magnetic ions can absorb the excitation
energy or trap photocarriers. However, as shown in
Figs.~\ref{fig:znmno_PL}(b) and~\ref{fig:znmno_PL}(d) the intensity of the
PL lines in both polarizations increases nonlinearly with the magnetic field. As a consequence, total PL intensity increases by an order of magnitude when magnetic field increases from 0 to 7~T, as shown in Fig.~\ref{fig:znmno_PL}(d). This indicates that applying a  magnetic field limits the impact of non-radiative processes on the emission from the excitonic states. A similar effect was reported for Zn$_{1\textrm{-}x}$Mn$_{x}$Se,\cite{Lee:2005_PRB} and explained by spin polarization of carriers and Mn$^{2+}$ ions, which blocked the transfer of the excitation to Mn ions.

The energy positions of the PL lines are depicted in
Fig.~\ref{fig:znmno_PL}(c) as a function of the magnetic field. The
strongest line exhibits a nonlinear redshift and $\sigma^+$ polarization. The redshift starts to saturate around $B=2$~T and attains 5~meV at 7~T. As
discussed in the previous section, a similar magnitude of the redshift is
found for the ground state exciton $A$ in the reflectivity measurements in
the same polarization. Interestingly, in addition to the PL line in the
$\sigma^+$ polarization, a somewhat weaker PL line emerges at {\em lower}
photon energies in the $\sigma^-$ polarization. Actually, in the previous
PL studies\cite{Przezdziecka:2006_SSC} the lowest  exciton line did also exhibit a redshift and the $\sigma^-$ polarization in the presence of a magnetic field.

As discussed in Sec.~\ref{Magneto-reflectivity}, the electron-hole exchange contributes significantly to the splitting between the $A$ and $B$ free excitons in ZnO. The influence of the electron-hole exchange is much weaker in the case of bound excitons than in the case of free excitons. Thus, the $A$-$B$ splitting is  much smaller in the case of bound excitons.\cite{Meyer:2010_PRB} Actually, we show below (next part and Fig. \ref{fig:znmno-integrals}) that the Zeeman shift of the bound excitons can be described assuming no contribution of the electron-hole exchange, as in the case of an acceptor-bound or donor bound exciton.  Moreover, the magnetic field induced splitting of the $B$ exciton, originating from the heavy-hole-like $\Gamma_9$ band, is larger than (and opposite in sign to) the splitting of the exciton $A$, which is associated with the $\Gamma_7$ band. The opposite circular polarization of the lowest line in the excitonic reflectivity and PL spectra observed at higher magnetic fields suggests therefore that a difference of binding energies of the $A$ and $B$ excitons is comparable to the relative energy distance between valence bands from which they originate. When $A$ and $B$ excitons separation energy at zero magnetic field is negligible, their relative order at magnetic field is governed by the difference of the respective shifts induced by the $s,p-d$ interaction. To sum up, at low magnetic fields, the $A$ and $B$ excitonic lines are too broad (FWHM of $\sim$20~meV) to be resolved. At higher magnetic fields, $\Gamma_7$-related exciton $A$ is observed in $\sigma^+$ polarization and $\Gamma_9$-related exciton $B$ is observed in $\sigma^-$ polarization.

Typically, thermalization processes lead to a stronger emission from a lower energy state. It is not the case of the present study: the higher energy PL component ($\sigma^+$) exhibits a stronger intensity. We explain this observation in terms of the mechanism of the bound exciton formation. Prior to any bound state formation photocreated holes relax to the lowest  $\Gamma_7$ valence band (order determined from reflectivity results for free carriers). Next, they form the corresponding bound excitons. This results in the strong $\sigma^+$ polarized emission. However, some excitons change their symmetry and transfer to a bound state involving a $\Gamma_9$ related hole, being in the magnetic field the lowest bound exciton energy state. Radiative recombination of these excitons results in $\sigma^-$ polarized emission. Its intensity is governed by the ratio of $\Gamma_7$ related exciton formation and recombination and $\Gamma_7$ to $\Gamma_9$ excitonic transfer rate. The stronger emission in $\sigma^+$ then in $\sigma^-$ polarization indicates that excitonic transfer rate from $\Gamma_7$ band to $\Gamma_9$ band is relatively small.

\section {Evaluation of the $p-d$ exchange integral}
\label{sec:discussion}

\begin{figure}
\centering
\includegraphics*[width=0.9\linewidth]{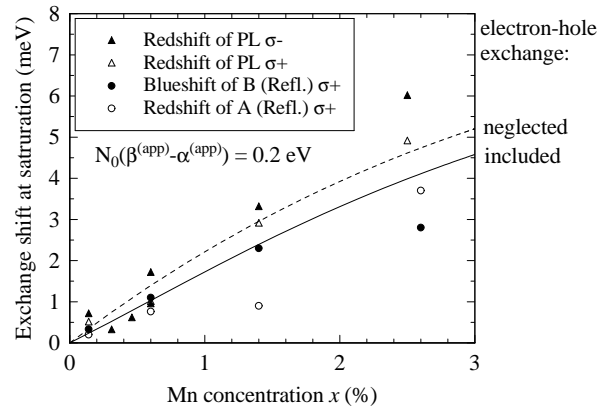}
\caption[Exciton shifts vs $x$ (Zn$_{1-x}$Mn$_{x}$O)]{Shift
of exciton energies observed in reflectivity and photoluminescence measurements on Zn$_{1-x}$Mn$_{x}$O plotted as a function of the Mn concentration $x$.
The shifts are determined at saturation of the Mn spin polarization: the
redshift of the exciton $A$ by using reflectivity in the
$\sigma^+$ polarization (empty  circles); the blueshift of the exciton $B$
in the $\sigma^+$ circular polarization by using reflectivity
(full circles); the redshift of the bound exciton from PL
measurements in the the $\sigma^- $ polarization (full triangles) and in the $\sigma^+ $ polarization (empty triangles). PL data for
three diluted samples are taken from
Ref.~\onlinecite{Przezdziecka:2006_SSC}. Solid line represents the value of
the shift calculated for the heavy-hole-like $\Gamma_9$ band assuming
${N_0(\beta^{\mathrm{(app)}}-\alpha^{\mathrm{(app)}})=+0.2}$~eV, and is
more suitable for the exciton $B$. Dashed line corresponds to the
calculation with the neglected electron-hole exchange interaction and is
more suitable for bound excitons seen in the PL, $\sigma^- $ polarization.} \label{fig:znmno-integrals}
\end{figure}

In order to evaluate the magnitude of the $p-d$ exchange energy for
Zn$_{1\textrm{-}x}$Mn$_{x}$O we have collected in
Fig.~\ref{fig:znmno-integrals} the determined redshifts of the exciton $A$
and the blueshifts of the exciton $B$ at magnetization saturation (low temperatures and high
magnetic fields), as a function of the Mn concentration. The redshifts
of the PL lines are also shown. Taking into account possible errors in the
energy and Mn content determination, the relative accuracy is of the order
of 20\% in the case of both axes. Theoretical
lines in Fig.~\ref{fig:znmno-integrals} represent the expected energy shift
for an exciton associated with the heavy-hole-like $\Gamma_9$ band for
which the shift is larger than for the one linked to $\Gamma_7$ band,
assuming $N_0(\beta^{\mathrm{(app)}}-\alpha^{\mathrm{(app)}})=+0.2$~eV, $S
= 5/2$, and $x_\textrm{eff}=x(1-x)^{12}$.
The solid (dashed) line is obtained taking into account (neglecting) the
electron-hole exchange interaction. As seen, the exchange shift determined from the reflectivity (the PL) measurements is reasonably well described by the solid (dashed) line. This is expected since the electron-hole interaction affects much less bound excitons which are observed in the PL.

Taking the determined value of $N_0(\beta^{\mathrm{(app)}}-
\alpha^{\mathrm{(app)}})=+0.2\pm0.1$~eV and assuming a standard value of
$N_0\alpha^{\mathrm{(app)}}=+0.3\pm0.1$~eV,\cite{Chanier:2009_PRB,Andrearczyk:2005_PRB,Beaulac:2010_PRB} we obtain the $p-d$ exchange energy
$N_0\beta^{\mathrm{(app)}}=+0.5\pm0.2$~eV.  If our assignment of the
valence band order were incorrect, the resulting value of the $N_0\beta^{\mathrm{(app)}}$ would be
$N_0\beta^{\mathrm{(app)}}=+0.1\pm0.2$~eV. We note that the latter value
includes zero and negative values within the uncertainty limits. We can
estimate the upper limit of the absolute value of the $p-d$ exchange
integral $|N_0\beta^{\mathrm{(app)}}|<0.7$~eV.

\section{Conclusions}

The giant Zeeman splitting of free excitons has been observed in
magneto-reflectivity experiment performed on Zn$_{1\textrm{-}x}$Mn$_{x}$O
samples with the Mn concentration $x$ ranging from 0.14\% to 2.6\%. In a
view of our findings, Zn$_{1\textrm{-}x}$Mn$_{x}$O appears in many aspects
to be a typical II-VI DMS: its magnetization is well described by the
paramagnetic Brillouin function, circular polarization of the giant Zeeman
effect and, thus, the sign of the MCD, are the same as in, {\em e.~g.},
Cd$_{1-x}$Mn$_{x}$Te. Although the incorporation of Mn degrades
optical quality of the crystal, excitonic transitions are
visible in reflectivity and photoluminescence even in the case of samples
with a relatively high concentration of magnetic ions. The most important
difference is that the overall amplitude of the exciton splitting is one
order of magnitude smaller than the one expected for the value of the $p-d$
exchange energy suggested by the Schrieffer-Wolf
theory,\cite{Blinowski:2002_MRS, Beaulac:2010_PRB} $N_0\beta = - 3.1 \pm 0.1$~eV  or implied by x-ray
spectroscopy, $N_0\beta = - 3.0$~eV.\cite{Okabayashi:2004_JAP}
By a quantitative interpretation of the reflectivity spectra, it has been
possible to find out how to link these puzzling results to a structure of
the valence band in ZnO and to the strong $p-d$ hybridization, expected for
materials with the short anion-cation distance.

More specifically, experimental results and their interpretation
presented here for Zn$_{1\textrm{-}x}$Mn$_{x}$O and earlier for
Zn$_{1\textrm{-}x}$Co$_{x}$O,\cite{Pacuski:2006_PRB}
Ga$_{1\textrm{-}x}$Mn$_{x}$N,\cite{Pacuski:2007_PRB,Suffczynski:2011_PRB}
and  Ga$_{1\textrm{-}x}$Fe$_{x}$N,\cite{Pacuski:2008_PRL} lead to several
conclusions concerning exciton magneto-spectroscopy and properties of wide
band-gap wurtzite DMSs.

First, a meaningful determination of exciton energies from
magneto-reflectivity spectra requires the application of the polariton
model that incorporates optical transitions to excitonic excited states and
band continuum.

Second, the determined free exciton energies as a function of the magnetic
field indicate that the giant Zeeman splitting is larger for the exciton
$B$ than for the exciton $A$ in (Zn,TM)O. This means that the exciton $B$
is associated with the heavy-hole-like $\Gamma_9$ valence band.  This is in
contrast to (Ga,TM)N, where a giant splitting is larger for the exciton
$A$, in agreement with the notion that the heavy-hole from the $\Gamma_9$
valence band contributes to the ground state exciton $A$ in GaN.

Third, quantitative description of excitonic splitting and selection rules requires the model containing the spin-orbit coupling as
well as the $s,p-d$ and $s-p$ exchange interactions.

Fourth, the obtained results point to the ferromagnetic character and the
small magnitude of the apparent $p-d$ exchange energy,
$N_0\beta^{\mathrm{(app)}}=+0.5\pm0.2$~eV, in
Zn$_{1\textrm{-}x}$Mn$_{x}$O. These findings, and similar ones for
Zn$_{1\textrm{-}x}$Co$_{x}$O,\cite{Pacuski:2006_PRB}
Ga$_{1\textrm{-}x}$Mn$_{x}$N,\cite{Pacuski:2007_PRB,Suffczynski:2011_PRB}
and  Ga$_{1\textrm{-}x}$Fe$_{x}$N,\cite{Pacuski:2008_PRL} substantiate the
model predicting the reversed sign and reduced magnitude of the splitting
of extended valence band states once the potential produced by individual
magnetic ions is strong enough to bind a hole.\cite{Dietl:2008_PRB}

Finally, we comment on consequences of our findings for the search for the
hole-mediated ferromagnetism in oxide and nitride DMSs. As already
noted,\cite{Dietl:2008_PRB} a significant contribution of the $p-d$
hybridization to the hole binding energy shifts the insulator-to-metal
transition to higher hole concentrations, making the conditions for the
hole delocalization, a necessary ingredient for the operation of the $p-d$
Zener mechanism of ferromagnetism, difficult to achieve.


\begin{acknowledgements}
The work was supported by the EC through the FunDMS Advanced Grant of the
European Research Council (FP7 "Ideas") and Polish public founds in years
2010-2013 (Polish MNiSW project Iuventus Plus and Polish NCBiR project LIDER). We thank Roman St\mbox{\c{e}}pniewski, Alberta Bonanni, and Hideo Ohno for valuable discussions at various stages of this work.
\end{acknowledgements}

\end{document}